\documentclass[11pt,twoside]{article}

\usepackage{asp2006}
\usepackage{epsf}

\def\hb{\ifmmode {\rm H}\beta \else H$\beta$\fi}
\def\civ{C\,{\sc iv}\,$\lambda1549$}
\def\Msun{\ifmmode M_{\odot} \else $M_{\odot}$\fi}
\def\Lsun{\ifmmode L_{\odot} \else $L_{\odot}$\fi}

\begin{document}
\title{Conference Summary}   
\author{Hagai Netzer$^1$ and Joseph C. Shields$^2$}   
\affil{$^1$School of Physics and Astronomy and the Wise Observatory, Tel Aviv University, Tel Aviv 69978, Israel}
\affil{$^2$Physics \& Astronomy Department, Ohio University, Athens, OH 45701 USA}    

\begin{abstract} 
The meeting in Xi'an covered the enormous and continuously expanding area of AGN research,
from theory to the most sophisticated observations and from $\gamma$-ray energies to long radio
wavelengths. The short summary below gives some, but definitely not all, highlights and new
results presented by the participants. 
\end{abstract}

\vspace{-0.1in}
\section{Black holes}
The main progress in this area in recent years, that enables comprehensive statistical
studies of active black holes (BHs), is the success of the large reverberation mapping
project. This allows reliable estimates of broad line region (BLR) sizes and BH masses.
The method was reviewed by S. Kaspi
and B. Peterson with additional (somewhat intriguing and perhaps conflicting) suggestions
 by M. Benzt. The talks summarized the
10 years long project and illustrated the accuracy of
the ``single epoch mass determination'' method. The original reverberation mapping 
work was based on the measurements of the optical
(5100\AA) continuum and the \hb\ line. It provided good  (factors 2--3 uncertainty) estimates
of BH masses at low redshifts, where \hb\ is accessible to ground-based spectroscopy.
The main concern and the biggest
unknown is the extension of the method to high redshifts where \hb\ measurements are no longer
available. While some workers suggest that the combination of the UV continuum and
the \civ\ line is a proper substitute for the \hb$-$L(5100\AA) method,
others quote factors of 3, 5 and even larger uncertainty on the masses obtained in this way.
A possible way to proceed is to use ground-based J,H and K spectroscopy to measure the \hb\ line in
high redshift sources. In principal, this could help solve most problems. In reality, the observations
are difficult and the number of AGNs measured so far quite small.

Notwithstanding those difficulties, we have seen several attempts in this meeting to explore
the land of BH  mass at both extremes. Very low mass BHs, like those presented by J. Greene,
are very useful indicators of the relationship between the host galaxy and the central BH at low
redshifts. Very high mass BHs, those approaching 10$^{10}$ \Msun\ and presented by F. Hamann,
are some of the best indicators
of the early evolution of massive galaxies. Additional pieces of information, related
perhaps to the metal enrichment in high redshift sources, are the correlation of BLR metallicity with
either BH accretion rate or BH mass.

Of the various other talks related to BHs we mention the very interesting
suggestions by A. Marconi and Ran Wang to construct the fundamental plane for active BHs.
Such methods have been very successful in studying various types of normal galaxies and can, perhaps, work also
for active galaxies. At this stage it is not very clear what the best parameters are. One suggestion
involves the X-ray and radio luminosities and the BH mass. This would make it difficult
to include radio-quiet AGNs in the same scheme. The idea needs more exploration and refinement.

\section{Accretion Disks and X-ray variability}

AGN accretion disks have been explored, in great detail, since the late 1980s. These
are complicated systems that defy most attempts to compute their emergent spectrum
even for static, non-disturbed disks. Needless to say, unstable disks, and those with strong
magnetic fields and  hot coronae, are even more complicated. Thus, it is no surprise that
progress in this area has been slow.

Some refreshing news on more realistic disk calculations
was reviewed by O. Blaes who introduced
a very detailed attempt to model disks on all scales. The new calculations  contain
several of the earlier missing ingredients.
 MHD effects are included and suggest that AGN
disks are probably supported by magnetic pressure, which is more important than the
gas and the radiation pressure.
The comparison of broad-band calculated spectra with observations of SDSS sources suggests
good agreement in the optical part of the spectrum but less agreement
with the UV part. Even specific spectroscopic features, like the Balmer jump, can
be computed, reliably, by the new codes.

Unfortunately, there is still a long way to go and there are  various components that are not
included in the calculations. A notable example is the absence of a hot corona which
made several of the X-ray participants  unhappy. We will have to wait a little longer before
such components are included and before we can answer the very interesting question of whether the soft
($< 1$\, keV) X-ray spectrum of AGNs, at least in those objects hosting the smaller BHs,
 is the short wavelength tail of the central accretion disk spectrum.

 X-ray variability is definitely an area of great excitement (physically as well as the reaction of
 the audience). We have heard several talks that summarized not only major discoveries of recent years
 but also the comprehensive analysis of the huge data base on AGN variability. Some of the highlights
 include
 \begin{description}
 \item[Narrow variable X-ray lines:] These were reviewed by J. Turner who showed that such 
 features are a common phenomenon. So far they have been observed in eight AGNs and seem to originate
 very close to the BH, in the same area of the disk thought to produce the (by now somewhat questionable)
 relativistic disk lines. The narrow lines change their velocity and intensity and seem to have extremely large
 equivalent widths, up to 100 eV in some cases.  Such strong components cannot be explained by any disk-corona models.
 They may be used, when better
 statistics are available, to infer and perhaps even measure  the BH spin.
 \item[Broad relativistic X-ray lines:] This is still an area of some debate.
While some workers believe
 that the phenomenon has been observed in almost all AGNs,
others maintain that most of those (perhaps
 all but two or three) are not real and the data are entirely consistent with narrow K$\alpha$ lines.
  The statistical evidence was summarized by T. Yaqoob who
 defended the view that such lines are very common and seen in 25--50\% of all sources.
 This view is definitely not shared by the entire X-ray AGN community.
  Perhaps the only safe conclusion is
 that the parameters needed to describe broad K$\alpha$ lines, in combination with the unknown shape of
  the 4--8 keV continuum, leaves too much freedom for the suggested models.
 \item[X-ray variability and BH and accretion disk physics:]
 P. Uttley reviewed the great progress in this area. Several long
 term monitoring campaigns are starting to pay off and very sophisticated time series 
(or power spectral densities - PDSs)
 analysis can be carried out.
 It seems that a break in the power-spectrum is a common feature but the analogy between BHs in X-ray
binary systems and in AGNs is more complicated than 
 claimed in earlier works.
  According to Uttley, the variability pattern can be used to distinguish among
 several physical models, such as accretion flows, jets, etc, as the origin of the observed variability. The first
 of those seems to be the best current explanation.
 There are several exciting new results some based on
 single sources and others on the sample properties.
 In one case, (NGC\,4051) there is clear indication for QPO type
 variability. For the entire sample included in the review, that contains small (X-ray binary) and large
(AGN) BHs, the break frequency in the PDS (still the most fundamental property) obeys
 $t_{\rm break} \propto M_{BH}\dot{M}^{-1}$ where $\dot{M}$ is the normalized accretion rate.
  \end{description}

 \section{Jets}
 Perhaps the most rapid progress in AGN observations in recent years is related
 to blazars and the very high energy observations of such sources.
 Most important perhaps is the big advance in atmospheric Cherenkov Telescope technology.
 We have seen beautiful observations of jets from the radio to X-ray, Gev and even Tev energies
 and details that were not available even a couple of years ago.
Some of the new results were presented by S. Wagner who argued that the
source of the observed $\gamma$-rays is inside the jet. He also argued that at least
in one source (Mkn\,421) there is some Tev emission.
There is also new evidence for low energy cut-offs and extremely compact emission regions,
of order of the Schwarzschild radius of the central BH, in some sources.
L. Maraschi showed us newly detected extremely large X-ray jets (courtesy of the {\it Chandra}
observatory)  confirming high electron temperatures many kpc away from the center.

 Unfortunately, in this area the models seem to be lagging behind, not because they
 are not sophisticated enough, some  are extremely complex, but because
 of the huge number of free parameters.
 Very interesting ideas have been presented by
 M. Boettcher who argued for the differentiation of hadron jets and lepton jets.
 We have learned that
 spectral information, by itself, is not enough to differentiate between all those possibilities
 and measurements of variability, preferably on very short (intra-day) time scales, are sorely needed.
 It seems that none of the available models can uniquely explain the new observations.
 In this respect, the very high energy land of AGN is yet another area of astronomy where
  observations lead the way.

Powerful jets, be it in the radio or the Gamma-ray, may well be related
to the structure and the stability of the central accretion disk. This was
explained in great detail by J.M. Wang who emphasized the very important
and likely physical correlation between the very small (innermost parts of
the accretion disk) and the very large (Mpc size jets) scales.

\section{Winds and outflows}

This area is progressing very rapidly, observationally and theoretically. Much of the progress is
due to several superb spectroscopic observations, in the ultraviolet, optical and X-ray bands.
Present day X-ray wind models can be compared to many dozens of X-ray absorption lines, in a handful
of sources, that indicate the motion of the ``warm'' ($\sim 10^5$\,K) gas. The number of UV lines
is not as large but their velocity can be investigated to a superb accuracy, a few km/sec.
An extremely nice result of this type was given by M. Crenshaw who showed the first example of a cool
component, in NGC\,4151,  that is seen both in emission and in absorption. There were other interesting
results related
to  X-ray observations. One example is the M. Guainazzi  sample of many Seyfert 2s that, despite having
only moderate S/N (XMM-Newton) observations, seem to be similar in most properties to NGC\,1068.

There are big discrepancies between various models (or explanations). The Guainazzi sample
of Seyfert 1s and 2s suggests that the X-ray absorbing gas seen in the first class of sources (the so-called
``warm absorber'') is the X-ray emitting gas seen in the second class. This would indicate that the 
size of the warm-absorber region
is tens and perhaps hundreds of pc. On the other hand, the F. Nicastro analysis
of the X-ray spectrum of NGC\,4051, suggests a sub-pc size of the warm absorber in this source. 
Is this unique to NGC\,4051 or, perhaps, there is a problem of interpretation?
Interestingly, the distance of the warm absorber in NGC\,3783 (perhaps the best studied case
so far) as deduced
from a detailed photoionization analysis, is roughly the geometrical mean of the two (about 1--3 pc).

A beautiful demonstration of the usefulness of the multi-wavelength approach was given by S. Gallagher who
reviewed BAL QSOs. Geometry must be very important in such cases. It determines the fractional amount
of UV and X-ray radiation that reach the observer and perhaps also the total width of the absorption
troughs. The combination of X-ray and UV observations provides a powerful tool to infer the viewing angle
to the center and the column density of the absorbing gas. A key observational result is that the X-ray
luminosity gets weaker when the line velocities are higher. It seems that the ratio of L(UV)/L(X-ray) in
those sources  can be explained by a high-mass  wind that drives
material almost in parallel to the central disk surface. So far we have been limited by the very small
number of sources needed for this analysis. However, the numbers are getting larger (35 in some of
Gallagher's diagrams) and a clearer
picture starts to emerge.

The theory developed to explain these observations is, again,
lagging behind. Much of what was
shown in the meeting regarding wind models was
 based on numerical modeling of photoionized gas which is streaming along radial or curving,
even spiraling stream lines. Despite a great body of observations,
those models are not unique. As explained by D. Proga,
this is not surprising given the huge parameter space in disk-wind models. It is not even clear whether
we are witnessing a smooth freely expanding wind or whether there is a  ballistic type
motion that requires confinement.

Theoretical arguments of a different type were presented by D. Chelouche who argued that thermal winds can explain
many of the observed properties. Such winds are likely to develop on large scales, between the
BLR and the NLR, and the typical velocity is consistent with a temperature of several million
degrees. Such hot outflows can carry with
them cooler gas that is responsible for much of the absorption. The speaker reminded us also of the importance of
high quality observations and the combination of UV and X-ray data. Apparently, this combination in
stars resulted in the suggestion
that previous mass outflow rate estimates were off (too high) by a factor 10 (!!!). Can this also be
the case in AGN outflows?

\section{The broad line region}

There are several important issues regarding the BLR geometry and gas composition
although none is very new. W. Kollatschny presented
the best case, so far, for measuring gravitational redshhift components in broad
emission line profiles (the He{\sc ii} $\lambda 4686$ line in Mkn\,110). This is a
tricky issue that requires a unique combination
of variability and velocity because the magnitude of the shift is
relatively small. There is clearly a need to look for lines that originate closer to
the black hole, where the effect is stronger. 
 Two dimensional transfer functions were also presented in the same review. They beautifully
 demonstrated the ionization and velocity stratification and prove, yet again, how useful and how time
 consuming the application of this  method is.

 The SDSS sample provides a great opportunity to investigate several BLR properties as a function of
 luminosity. Some new results were presented by Nagao referring mainly to the gas metallicity and using
 grouped rather than individual sources. Apparently,
 metallicity is luminosity (but not redshift) dependent. Here, again, it seems difficult to decide whether
 the luminosity or the accretion rate are driving the observed correlations.

 Dust is an important ingredient in all regions, from the outskirts of the BLR to well inside
 the host galaxy. The implication to the BLR size were reviewed by A. Laor as discussed in the
 next section.

Finally, it was nice to be reminded that some of the main puzzles of
BLR research are still with us, after 20 or more years of study. One such
example was discussed by M. Joly who pointed out that collisional (rather
than photoionized) gas may be a necessary ingredient to explain the strong
observed FeII lines. Another outstanding question is the so-called
``Baldwin effect'' that was reviewed by J. Shields. There are several suggested
explanations for this line-to-continuum correlation. Is the correct explanation 
somewhere in that list?
We probably need more time, and perhaps more data to answer this question.

\section{Dust}   

The role of dust in AGNs was a major theme at this meeting.  It is
clear more than ever that a detailed understanding of the effects of
dust is essential to progress in the field.  The influence of dust
emerged in several lines of discussion.

Dust is central to our understanding of the {\em torus}.  Recent
observations described at the meeting indicate that this structure is
smaller than originally thought, but resolved in at least some nearby
systems.  NGC~1068 remains a favorite object of study; for this
source, W. Jaffe showed $8 - 13.5 \mu$m interferometric measurements
that reveal a source with an extent of $\sim 2 \times 3$ pc that can
be identified with the torus.  For the same object, R. Davies
presented infrared H$_2$ detections that likely trace the same
structure on somewhat larger scales.

In addition to being small, the torus is likely to be clumpy.  As
reviewed by M. Elitzur, a clumpy structure makes it possible to
reconcile the infrared spectral energy distribution and the implied
(large) range of grain temperatures with the torus's small radial
extent.  The application of these ideas to the specific case of
NGC~1068 was presented by S. H{\"o}nig, based on 3D radiative
transfer simulations.  In this picture, a single clump can be highly
optically thick, allowing a large gradient in grain temperature to
exist between the irradiated and shadowed sides of the cloud.

Dust remains essential to understanding the Seyfert 1/2 dichotomy.
The dust sublimation radius evidently defines the outer boundary of
the broad-line region (BLR); A. Laor reviewed the theoretical basis
for this statement, and M. Suganuma presented the current status of
reverberation mapping for the infrared continuum versus UV/optical
line emission in Seyfert nuclei.  The results consistently place the
infrared-emitting medium just beyond the BLR, for sources spanning a
range of luminosity and hence radial scaling.  The infrared emitter
presumably corresponds to the torus, which causes us to observe a
Seyfert 2 nucleus when our line of sight to the BLR is obscured by the
dusty medium.  However, as noted by Elitzur, the fact that the
obscuring structure is lumpy rather than a homogeneous donut means
that obscuration and classification of an AGN as Type 1 or 2 becomes a
probabilistic function of inclination rather than a strict function of
viewing angle.

While this picture of geometric unification is clearly applicable to
many sources, the possibility remains that some Type 2 AGNs simply
lack a BLR.  Laor noted that in low-luminosity systems the dust
sublimation radius may move in to sufficiently small radii that the
(dust-free) BLR disappears.  Alternatively, Elitzur sketched a
scenario in which the BLR can be associated with a wind emerging from
the accretion disk, and posited that this outflow disappears at low
luminosity.  

An important tool for discovering hidden broad-line
regions (HBLRs) in apparent Seyfert 2s has been spectropolarimetry.
E. Moran reviewed the status of spectropolarimetric surveys and noted
that much of the work to date is characterized by strong selection
effects and sensitivity biases.  As a result it is difficult to draw
meaningful conclusions when comparing properties of objects with and
without detected HBLRs, which may be relevant to the question of
whether Seyfert 2s lacking a BLR exist.  Surveys are currently
underway to improve on this situation, but in interpreting the
spectropolarimetric results it is important to recognize that absence
of evidence does not constitute evidence of absence, since the
discovery of a hidden BLR by this method remains dependent on the
presence of a suitable scattering medium.  

Where dust survives near the AGN, it may play a fundamental role in
influencing the state of gas irradiated by the central source.
B. Groves reviewed models for the narrow-line region (NLR), and noted
that radiation-pressure-dominated dusty clouds likely provide the best
physical framework for understanding the tuning of parameters implied
by the observed emission-line ratios.  In specific instances, however,
the structure and energetics of the NLR may be dominated by other
phenomena.  N.  Bennert showed examples in which circumnuclear star
formation turns out to be a significant contributor to apparent NLR
emission; J. Holt and K. Inslip provided case studies in which AGN
feedback that includes radio jets gives rise to large-scale outflows in
the NLR.

A final important aspect of dust associated with AGNs concerns
high-redshift quasars and the evolutionary state of their host
galaxies.  The degree to which star formation precedes the onset of
quasar activity at early epochs is of great interest for understanding
the co-evolution of black holes and their surrounding stellar systems.
F. Hamann reviewed abundance constraints relevant to this topic; an
increasing number of observations point to large far-infrared
luminosities in many high-$z$ quasars, which in turn necessitate a
minimum mass of dust formed from heavy elements produced by stellar
nucleosynthesis that preceded the quasar activity we now observe.  The
implied amounts of star formation can be large and indicate that
galaxy-sized stellar systems were already in place before the quasar
turned on, in many cases.

\section{Probing Dust}

A growing number of strategies for probing the properties and influence
of dust in AGNs were discussed at the conference.  These include

\begin{description}

\item[IR Spectroscopic Features:]  A. Lee, E. Sturm, L. Hao,
  R. Deo, and others provided illustrations of the power of infrared
  spectroscopy for probing the physical properties of dust, its
  geometrical distribution, and its role as a reprocessor in AGNs.
  Of particular interest are the $\sim 10$ and $\sim 18 \mu$m silicate ``humps''
  and several PAH emission features.
 The impact of the {\em Spitzer
    Space Telescope} on this field has been significant, and was one
  of the highlights of this meeting.

\item[XAFS:] Methods usually associated with condensed matter
  experiments in terrestrial labs may see increasing application to
  astronomical dust in the future.  As described by J. C. Lee, a
  promising example is X-ray absorption fine structure (XAFS)
  analysis.  Observation of XAFS features with next generation X-ray
  observatories could provide a tool for measuring the composition of
  grains when a strong background X-ray source is available, as in an
  AGN.

\item[Extinction Law:]  The wavelength dependence of extinction
  contains information on the physical properties of the scattering
  and absorbing grains, and considerable effort has been invested to
  use the extinction law to learn about dust in the local interstellar
  medium.  The situation is more complex in AGNs, where the
  unattenuated SED in a given source is not well known and hence
  measurement of the extinction law itself is less secure.  Studies to
  date provide suggestive evidence that the grain size distribution in
  AGNs may be quite different from that in the local ISM, although
  these results have proven controversial.  At this conference
  B. Czerny reviewed the state of our understanding of the extinction
  law in AGNs based on analysis of AGN ensemble SEDs.  She presented a
  cautionary exercise in which the multi-epoch spectra of a single
  variable source, Mrk 335, were analyzed using the method appropriate
  for ensemble analysis, which assumes that SED variations stem from
  differences in the amplitude of extinction; the Mrk 335 analysis
  produced an extinction law quite similar to the ensemble result,
  even though the SED variations among its spectra probably have
  little to do with variable extinction!  There is clear evidence that
  the grains in AGNs differ from the local ISM (e.g. AGNs lack a 2175
  \AA\ extinction feature), but more effort will be needed to
  definitively measure the extinction law and draw detailed
  conclusions.

\item[Unification Tests:]  For cases where geometric unification
  seems to be a good bet (i.e. differences stem from viewing angle), a
  comparison of the properties of Type 1 and Type 2 objects provides
  constraints on the optical depth and spatial distribution of dust.
  At this meeting, K. Cleary presented the results of a study
  comparing 3C quasars and radio galaxies where  {\em Spitzer} observations
  make it possible to separate thermal and synchrotron contributions.
  The two populations show consistent luminosities, validating the
  unification assumption, after correction is made for nonthermal
  emission in the quasars and absorption in the radio galaxies; for
  the latter sources, the results directly yield the infrared optical
  depth for the obscuring medium.  B. Schulz presented related results
  for infrared fine-structure lines and silicate features in the 3CR
  sources, which validate the unification picture while placing
  constraints on the torus dust properties.

\end{description}

\section{Beyond Dust}

\subsection{The X-ray Absorbing Medium}

Several aspects of AGNs discussed at the meeting that might seem
plausibly linked to dust in fact probably require other explanations.
A dusty medium identified with the torus is responsible for optical
obscuration, but as reviewed by R. Maiolino, an additional component
of dust-free gas is required to explain the X-ray absorption observed
in at least some sources.  The clearest examples are those that
display X-ray absorption indicative of large column densities, but
nonetheless feature broad UV/optical lines.  Likewise, the total $N_H$
inferred from X-ray absorption is often greater than that implied by the
dust optical depth at 10 $\mu$m.  Variability in the absorbing column
on short timescales suggests that the high-$N_H$ X-ray absorber
characteristically resides at radii comparable to that of the
broad-line region; a particularly striking case study was described by
G. Risaliti, who showed a sequence of X-ray measurements for NGC~1365
interpreted as a change from Compton-thin to Compton-thick absorption,
and a return to Compton-thin conditions, over a total span of four
days.  As noted by Maiolino, the implied size scale in such cases is
smaller than the dusty medium resolved in mid-IR continuum emission.
In Elitzur's theoretical picture the dominant X-ray absorber, along
with the BLR, is identified with the dust-free inner part of a wind
emanating from the accretion disk.

There is a need as well as opportunities for improving our
understanding of the X-ray absorbing medium.  Additional investigation
of the time variability of absorption in individual sources would
clearly be valuable for identifying the location and structure of the
absorber.  In the study of AGN ensembles, Maiolino argued that our
knowledge of X-ray absorber phenomenology remains severely limited by
selection biases that cause highly absorbed sources to be
under-represented in typical samples.  The X-ray Background continues
to be an important constraint on the total number of absorbed sources
as a function of redshift.  Maiolino cautioned that some objects with
reflection-dominated spectra may not be Compton-thick but rather cases
where the central X-ray source has shut down recently and the
reflecting regions remain visible due to light travel-time delays.  In
principle such examples of ``fossil reflection'' provide another means
of probing the extent and coverage of high column-density gas in the
nuclear environment.

\subsection{Seyferts versus LINERs}

As mentioned above, radiation pressure acting on dust in the NLR may
be important in tuning the ionization and other properties of the
emitting plasma, but it seems that some additional parameter is
involved in determining the emission spectrum of a given source, if we
consider LINERs as well as Seyferts.  In the past it was possible to
consider LINERs and Seyferts as forming a continuous distribution
spanning the AGN side of the usual line-ratio classification diagrams.
This picture has changed as a result of measurements of very large
spectroscopic samples of galaxies from the Sloan Digital Sky Survey.
As discussed by Groves, the distribution of AGNs in the line-ratio
diagrams is genuinely {\em bimodal}, implying that the LINER/Seyfert
distinction has some real physical meaning and is not simply an
arbitrary division.

The separation in properties may  arise from  differences
in the ionizing SEDs for the two classes of object, such that LINERs
lack the optical/UV ``big blue bump'' characteristic of Seyfert nuclei
and quasars.
 Analyses using SDSS
data by A. Constantin and by J. Greene indicate that at low redshift,
LINERs and Seyferts of similar luminosity have characteristically
different black hole masses, with LINERs typically having larger
$M_{BH}$.  A consistency can then be drawn if the correspondingly
lower $L/L_{Edd}$ for the LINERs results in a different accretion
structure (e.g. an advection-dominated accretion flow or other
radiatively inefficient accretion flow) producing a harder continuum
than in the Seyferts.  At this conference I. Marquez presented new
X-ray observations of LINERs, confirming earlier indications that
nuclear point sources are often, but not always, evident.  There is
some confusion in the literature stemming from the fact that extended
nebular sources associated with infrared-luminous starbursts sometimes
show emission-line ratios formally consistent with a LINER
classification, and are referred to as such, although their energy
source and physical nature are probably very different than in truly
nuclear sources.  D. Rupke presented new {\em Spitzer} measurements of
fine-structure emission lines for both categories of sources; the
infrared lines allow a clearer differentiation of the two
types of LINER emitters, compared with the optical line ratios.

\section{Other New Results}

A variety of other new and noteworthy results were presented at the
meeting.  A list of some of the findings that caught our attention is
given here; this summary is in no way comprehensive.

\begin{description}

\item[New composite QSO spectra:] Z. H. Shang and E. Sturm presented new
  composite spectra which give a nearly continuous coverage from $\sim 0.1 -
  40$ $\mu$m, which takes advantage of recent {\em Spitzer}
  measurements.  The result beautifully illustrates the detailed continuum
  structure through this bandpass, which the current paradigm interprets
  as a combination of emission from the accretion disk and
  circumnuclear dust.

\item[A clustering constraint on AGN companions:]  The extent to
  which AGN activity is related to interaction with companion galaxies
  has been the subject of controversy for some time.  At this
  conference C. Li presented the 2-point cross-correlation function
  for an SDSS sample that includes a large number of narrow-line AGNs.
  This analysis finds a small excess of companions at separations of
  $< 70$ kpc, such that $\sim 1$\% of AGNs have an additional neighbor
  beyond that expected for normal galaxies.  The excess increases at
  larger accretion rate, as measured by $L$([O~{\sc iii}])/$M_{BH}$
  (where $M_{BH}$ is estimated from the $M_{BH}-\sigma$
  relation); however, if substantial numbers of AGNs are linked to
  close interactions, the resulting mergers are already complete in
  most cases, such that a companion is no longer evident.

\item[High-velocity HI wings as a tracer of feedback:]  The
  influence of AGNs on their surrounding galaxy and beyond is a
  critical issue in understanding galaxy evolution.  R. Morganti
  presented observations of 21-cm H~{\sc i} absorption in radio
  galaxies which reveal outflows with speeds of up to $\sim 1000$ km
  s$^{-1}$, presumably originating in interactions between radio jets
  and the surrounding medium.  The fact that the detected gas remains
  {\em neutral} while being accelerated to such velocities is
  remarkable.  The total mass and kinetic energy participating in such
  an outflow may be significant, and in turn can be expected to have
  nontrivial consequences for the host galaxy and surrounding
  intergalactic medium.

\item[Star formation indicators in the presence of an AGN:]
  N. Levenson reviewed the available tools for distinguishing the
  contributions from an AGN and from young stars in a given object.
  She noted that caution is often warranted; an example is the
  far-infrared continuum, where the emergent SED is established to a
  significant degree by the geometry of the dust distribution, rather
  than simply the energy source.  An AGN can also modify the
  observable properties of star-forming regions such that they appear
  different than in systems lacking an accretion source.  An
  informative example was presented by R. Mason for NGC~1097.  This
  galaxy displays a young nuclear star cluster as well as a weak AGN;
  both are energetically significant, but the PAH emission that would
  be expected to be excited by the cluster is absent, presumably due
  to PAH destruction by the AGN's radiation field.  

\item[The Stellar Initial Mass Function in AGNs:]  S. Nayakshin
  reviewed the observational evidence for a top-heavy IMF in the
  Galactic Center and presented arguments as to why this might be
  expected for star formation that occurs in disks on small scales in
  active nuclei.  An intriguing consistency was reported by M. Sarzi,
  from an analysis of {\em Hubble Space Telescope} spectra for the
  central $\sim 10$ pc of a sample of Seyfert 2 galaxies.  Those objects
  show evidence for young stars in some instances, but the majority of
  starlight seen traces old (many Gyr) stars, and intermediate-age
  populations are conspicuously absent.  If the IMF is indeed skewed
  to high masses, this could explain a lack of observable signatures
  of star formation for ages beyond $\sim 10^8$ yr, consistent with
  what is seen.

\item[New metallicity diagnostics for the NLR:]  Accurate
  measurement of metallicity in the NLR is a long-standing problem,
  and it remains challenging to separate out the influence of
  abundance from other thermal and ionization effects in determining
  emission-line strengths.  T. Nagao described strategies for
  constraining NLR metallicity from rest frame UV emission features,
  which is of particular interest for high-redshift sources in which
  such features are observed at optical wavelengths.  One option that
  shows considerable promise is to use a combination of the C~{\sc
    iv}/He~{\sc ii} and C~{\sc iii}]/C~{\sc iv} line ratios; the
  former is sensitive in part to abundance while the latter is
  sensitive to ionization parameter, allowing separation of the two.

\item[Deviations from $M_{BH}-\sigma$ at moderate redshift:]  The
  correlation between black hole mass and stellar velocity dispersion
  for the host galaxy has generated enormous interest as a clue to
  processes that link the growth and evolution of an AGN and the
  surrounding stellar body.  A full understanding of this relation
  requires that we measure the correlation as a function of time and
  not simply at the present epoch, in order to understand just how
  closely nucleus and host are locked in their evolution.  M. Malkan
  reported results of a study for Seyfert 1 galaxies at $z=0.36$, with
  $M_{BH}$ estimates derived from emission-line measurements.  The
  results indicate that black hole masses at a given stellar velocity
  dispersion are larger on average by a factor of $\sim 4$ in these
  objects, compared with what is found locally.  Preliminary analysis
  of {\em HST} imaging provides supporting evidence for this
  conclusion, in that the bulge luminosities are low compared to what
  would be expected from the $M_{BH}$ estimates.

\item[New candidates for tidal disruption of stars:]  Stars
  wandering too close to a black hole are headed for trouble.  While
  theoretical predictions for the result have been available for some
  time, credible candidates for this phenomenon have only appeared
  rather recently.  S. Gezari presented remarkable results for a
  flaring event in the nucleus of an inactive galaxy that was observed
  at multiple epochs with multi-wavelength coverage.  The results are
  consistent with a tidal disruption event, and from the measured SED
  and time evolution it is possible to constrain the black hole mass;
  a constraint on the black hole's spin may also be possible, since
  the spin determines the location of the innermost stable orbit and
  hence the inner extent of a transient accretion disk.  Future
  studies of this type will be facilitated with the new capability for
  large-scale temporal sampling of normal galaxies provided by
  Pan-STARRS and LSST.

\end{description}

\end{document}